
\input phyzzx
\pubnum{KEK--TH--306, KEK preprint 91--130, OCHA--PP-21}
\titlepage
\def\pref#1{\rlap.\attach{#1}}
\def\cref#1{\rlap,\attach{#1}}
\def\ie{{\it i.e.}}
\def\etc{{\it etc.}}
\def\tanh{{\rm tanh}}
\def\coth{{\rm coth}}
\def\cosh{{\rm cosh}}
\def\sinh{{\rm sinh}}
\def\ln{{\rm ln}}
\def\det{{\rm det}}
\def\ln{{\rm ln}}
\def\G{${\cal G}$}
\def\F{${\cal F}$}

\title{(Super-) String in Two Dimensional Black Hole and
Target Space Dualities
\foot{Talk at YITP workshop, \lq\lq Developments in Strings
and Field Theories'', Kyoto, JAPAN}
}
\author{Shin'ichi Nojiri\foot{e-mail address
: NOJIRI@JPNKEKVX, NOJIRI@JPNKEKVM}}

\vskip 1cm

\centerline{\it Theory Group, National Laboratory for
High Energy Physics (KEK)}
\centerline{\it Oho 1-1, Tsukuba-shi, Ibaraki-ken 305, JAPAN}

\vskip 0.5cm

\centerline{and}

\vskip 0.5cm

\centerline{\it Faculty of Science, Department of Physics}
\centerline{\it Ochanomizu University}
\centerline{\it 1-1 Otsuka 2, Bunkyo-ku, Tokyo 112, JAPAN}

\vskip 1cm

\abstract{
We review the recently proposed
string theory in two dimensional black hole background.
Especially, the structure of the duality in the
target space is discussed.
The duality is analogous to \lq\lq $R \rightarrow 1/R$" symmetry
of a compactified boson.
We consider the duality
in more general target space manifolds which have Killing symmetries
and we give an explicit formula which connects two different
manifolds which are dual to each other.
Superstring theory in two dimensional black hole
background is also discussed
based on supersymmetric $SU(1,1)/U(1)$ gauged Wess-Zumino-Witten
model.
}
\endpage

\REF\ri{E. Witten\journal Phys.Rev.&D44 (91) 314}
\REF\rii{R. Dijkgraaf, H. Verlinde and E. Verlinde, preprint PUPT-1252,
IAASNS-HEP-91/22}
\REF\rrxiv{A. Giveon, \lq\lq Target Space Duality and Stringy
Black Holes'', LBL-30671}
\REF\rrvii{E.B. Kiritsis, \lq\lq Duality in Gauged WZW Models'',
LBL-30747, UCB-PTH-91/21}
\REF\rraix{N. Sakai and I. Senda\journal Prog.Theor.Phys. &75 (86) 692}
\REF\rxii{S.B. Giddings and A. Strominger, preprint UCSBTH-91-35}
\REF\rxi{G.T. Horowitz and A. Strominger\journal Nucl.Phys.
&B360 (91) 197}
\REF\rrxix{S. Nojiri, \lq\lq Superstring in Two Dimensional Black
Hole'', FERMILAB-PUB-91/230-T, KEK PREPRINT 91-90, KEK-TH388,
OCHA-PP-18}
\REF\rv{M.M. Nojiri and S. Nojiri\journal Prog.Theor.Phys. &B83 (90)677}
\REF\riv{Y. Kazama and H. Suzuki\journal Nucl.Phys. &B321 (89)232}
\REF\riii{L.J. Dixon, J. Lykken and M.E. Peskin\journal
Nucl.Phys.&B325 (89)326}
\REF\rvi{E. Marinari and G. Parisi\journal Phys.Lett. &B240 (90)561}
\REF\rpi{T.H. Buscher\journal Phys.Lett. &B201 (88)466}
\REF\rpii{T.H. Buscher\journal Phys.Lett. &B159 (85)127
\journal Phys.Lett. &B194 (87)59}
\REF\rpiii{A.A. Tseytlin\journal Mod.Phys.Lett. &A6 (91)1721}
\REF\rraii{F. David\journal Mod.Phys.Lett. &A3 (88) 1651}
\REF\rraiii{J. Distler and H. Kawai\journal Nucl.Phys. &B321 (89) 509}
\REF\rravi{E. Brezin, V. Kazakov and A. Zamolodchikov\journal
Nucl.Phys. &B338(90)673}
\REF\rravii{D. Gross and N. Miljkovic\journal Phys.Lett. &B238 (90) 217}
\REF\rraviii{P. Ginsparg and J. Zinn-Justin\journal Phys.Lett. &B240
(90) 333}
\REF\rraiv{J. Polchinki\journal Nucl.Phys. &B346 (90) 253}
\REF\rrav{S. Das and A. Jevicki\journal Mod.Phys.Lett. &A5 (90) 1639}
\REF\rrai{G. Mandal, A.M. Sengupta and S.R. Wadia\journal Mod.Phys.Lett.
&A6 (91) 1685}
\REF\rrxviii{S.P. de Alwis and J. Lykken, \lq\lq 2d Gravity and the Black
Hole Solution in 2d Critical String Theory'', FERMILAB-PUB-91/198-T,
COLO-HEP-258}
\REF\rrxv{M. Bershadsky and D. Kutasov, \lq\lq Comment on Gauged
WZW Theory'', PUPT-1261, HUTP-91/A024}
\REF\rrxvii{E.J. Martinec and S.L. Shatashvili, \lq\lq Black Hole
Physics and Liouville Theory'', EFI-91-22}
\REF\rrrv{J. Distler and P. Nelson, \lq\lq New Discrete States of
Strings Near a Black Hole'', PUPT-91-1242}
\REF\rix{P. Di Vecchia, V.G. Knizhnik, J.L. Petersen
and P. Rossi\journal Nucl.Phys. &B253 (85)701}
\REF\rxviii{D.Z. Freedman and P.K. Townsend\journal Nucl.Phys.
&B177 (81) 282}
\REF\rxv{T. Kugo and I. Ojima\journal Phys.Lett. &B73 (78) 459
\journal Prog.Theor.Phys. Supp. &66 (79) 1}
\REF\rxvi{M. Ito, T. Morozumi, S. Nojiri and S. Uehara\journal
Prog.Theor.Phys. &75 (86) 934}
\REF\rxvii{N. Ohta\journal Phys.Rev. &D33 (86) 1681 \journal Phys.Lett.
&B179 (86) 347}
\REF\rvii{S. Nojiri\journal Phys.Lett. &B252(90)561
\journal Phys.Lett. &B253 (91)
63 \journal Prog. Theor. Phys. &85 (91) 671}
\REF\rviii{S. Nojiri\journal Phys.Lett. &B262 (91)419
\journal Phys.Lett.&B264 (91)57}
\REF\rxiii{N. Ishibashi, M. Li and A.R. Steif,
preprint UCSBTH-91-28 (Revised Version)}
\REF\rxiv{J.H. Horne and G.T. Horowitz, preprint UCSBTH-91-39}
\REF\rx{P. Di Vecchia, J.L. Petersen and H.B. Zheng\journal Phys.Lett.
&B174 (86)280}

\chapter{Introduction}
Recently it was shown that the $SU(1,1)/U(1)$ gauged
Wess-Zumino-Witten
(GWZW) model describes strings in a two dimensional black
hole\pref\ri
This model gives a simple toy model of black hole, \ie ,
an example of space-time which has singularities.
The string propagation and Hawking radiation in this black hole were
discussed in Ref.\rii .
Since this model is a model of string,
the target space has a structure of duality\cref{\rii,\rrxiv,\rrvii}
which is analogous to \lq\lq $R \rightarrow 1/R$" symmetry
of a compactified boson\pref\rraix Furthermore,
this model can be regarded as the two dimensional gravity coupled
with $c=1$ conformal matter (free boson)
when the level $k$ of $SU(1,1)$ current algebra equals to $9/4$
($c=26$).
We expect that
this model could be one of toy models which provide a clue to solve
the dynamics of more \lq\lq realistic" string models.

The supersymmetric extension of this model appeared\refmark\rxii
as an exact solution of
ten-dimensional superstring theory corresponding to black
fivebranes\pref\rxi
In Ref.\rrxix , the supersymmetric extension
based on $SU(1,1)/U(1)$ supersymmetric GWZW (SGWZW) model
was discussed.
It has been shown\refmark\rv that supersymmetric $SU(1,1)/U(1)$
coset model has $N=2$
supersymmetry due to Kazama-Suzuki\refmark\riv mechanism and
this model is equivalent to $N=2$ superconformal models proposed
by Dixon, Lykken and Peskin\pref\riii
The central charge $c$ of this system is given by,
$$c={3k \over k-2} \ . \eqn\abi$$
Here $k$ is the level of $SU(1,1)$ current algebra.
When $k={5 \over 2}$, the central charge $c$ equals to 15 and
this conformal field theory describes a critical
Neveu-Schwarz-Ramond string theory.
The $N=1$ supergravity coupled with $c={3 \over 2}$
($\hat c=1$) superconformal matter would be described by
this critical theory.
Furthermore the $N=2$ superconformal symmetry of this model suggests
that pure $N=2$ supergravity would be also described by this model
when $c=6$ ($k=4$).
Due to $N=2$ superconformal symmetry, superstring theories in the two
dimensional black hole background can be constructed by imposing GSO
projection.
We expect that this superstring theory would be a continuum theory
corresponding to Marinari and Parisi's superstring theory in
one dimension\refmark\rvi.

In the next section, we review the relation between $SU(1,1)/U(1)$
GWZW model and string theories in two
dimensional black hole based on Ref.\ri .
In section 3, the duality structure of the target space is discussed.
We also consider the duality
of more general manifolds which have Killing symmetries.
We derive a formula which appeared in Ref.\rpi .
This formula connects two different
manifolds which are dual to each other.
In section 4, we briefly review the relation between $SU(1,1)/U(1)$
GWZW model and two dimensional gravity.
In section 5, the supersymmetric extension of this model is discussed
in the basis of $SU(1,1)/U(1)$ GWZW model
And the last section is devoted to summary and discussion.

\chapter{$SU(1,1)/U(1)$
gauged Wess-Zumino-Witten model and string theories in two
dimensional black hole}

The action $I^{\rm WZW}(g)$ of Wess-Zumino-Witten (WZW) model
is given by,
$$\eqalign{I^{\rm WZW}(g)
=&{k \over 8\pi}\int d^2x {\rm tr}g^{-1}\partial_\mu g
g^{-1}\partial^\mu g
\cr
&-{ik \over 12\pi}\int d^3x \epsilon^{\mu\nu\rho}
{\rm tr}g^{-1}\partial_\mu gg^{-1}\partial_\nu g
g^{-1}\partial_\rho g \ . }\eqn\ti$$
Here a matrix field $g$ is an element of a group $G$.
The Lagrangian theory corresponding to the $G/H$ coset model
($H\subset G$) is given by gauging a vector symmetry,
$g \rightarrow h^{-1}gh$ ($h\in H$), in the action \ti .
The action $I^V(g,h_{L,R})$ of the GWZW model takes the following form:
$$I^V(g,h_{L,R})=I^{\rm WZW}(h_L^{-1}gh_R)-I^{\rm WZW}(h_L^{-1}h_R)
\ . \eqn\tii$$
Here matrix fields $h_L$ and $h_R$ are elements of $H$.
If $H$ is an abelian subgroup, we can gauge an axial symmetry,
$g \rightarrow hgh$, instead of the vector symmetry.
The corresponding action
$I^A(g,h_{L,R})$ is given by
$$I^A(g,h_{L,R})=I^{\rm WZW}(h_Lgh_R)-I^{\rm WZW}(h_L^{-1}h_R)
\ . \eqn\tiia$$
Since we now consider $SU(1,1)/U(1)$ GWZW model, the gauge symmetry
can be vector \tii\ or axial \tiia .
The action in the vectorially-gauged WZW model \tii\ has the form
different from that of the action in the axially-gauged WZW
model \tiia\ but the corresponding conformal field theories are
equivalent. As we find in the next section, these two models are
related to each other by the duality.

There are compact and non-compact abelian subgroups in $SU(1,1)$.
When we gauge the compact subgroup, the resulting model describes
the string propagation in a two dimensional Euclidean manifold.
On the other hand, the target space which has Minkowski signature
is obtained by gauging the non-compact subgroup.

For a while, we consider the case that the axial symmetry generated
by the compact abelian subgroup of $SU(1,1)$ is gauged.

By parametrizing $g$ by,
$$g=\exp ({i \over 2}\phi_L\sigma_2) \exp ({1 \over 2}r\sigma_1)
\exp ({i \over 2}\phi_R\sigma_2) , \eqn\tiii$$
with $\sigma_i$ the Pauli matrices,
we obtain the action $I^{SU(1,1)/U(1)}$
of the $SU(1,1)/U(1)$ GWZW model\foot{
By using the Polyakov-Wiegmann formula, the action is given
by the integration of Lagrangian density in two dimensions since
the Wess-Zumino term is a total derivative in the parametrization
\tiii .}
$$\eqalign{I^{SU(1,1)/U(1)}=&{k \over 4\pi}\int d^2z
[-\partial \phi_L\bar\partial \phi_L
-\partial \phi_R\bar\partial \phi_R \cr
&-2\,{\rm cosh}\,r\,\partial\phi_L\bar\partial\phi_R
+\partial r\bar\partial r]\cr
&+{k \over 2\pi}\int d^2z [-(1+{\rm cosh}\,r)A\bar A \cr
&+A(\bar\partial \phi_R+{\rm cosh}\,r\,\bar\partial \phi_L)
+(\partial \phi_L
+{\rm cosh}r\, \partial \phi_R)\bar A] \ .}\eqn\tiv$$
Here gauge fields $A$ and $\bar A$ are defined by,
$$A\equiv h_L^{-1}\partial h_L ,
\ \ \ \ \bar A\equiv h_R^{-1}\bar\partial h_R\ . \eqn\tv$$
The action $I^{SU(1,1)/U(1)}$ is invariant under the following
axial gauge transformation:
$$\delta\phi_{L,R}=\alpha \ , \ \
\delta A=-{i \over 2}\partial\alpha \ , \ \
\delta \bar A=-{i \over 2}\bar\partial\alpha \ .\eqn\tvi$$
We fix the gauge symmetry \tvi\ by choosing the following gauge
condition:
$$\phi_L=-\phi_R=\phi \ .\eqn\tvii$$
By integrating gauge fields $A$ and $\bar A$ and rescaling
$r \rightarrow {r \over 2}$, we obtain,
$$I={k \over \pi}\int d^2z(\partial r\bar\partial r+\tanh^2r
\partial \phi \bar\partial \phi)
-{1 \over \pi}\int d^2z \sqrt h R^{(2)}[2\,\ln (\cosh\,r+1)+c]
\ .\eqn\tviii$$
Here $h=\det\,h_{\mu\nu}$ ($h_{\mu\nu}$ is a metric on the
string world sheet), $R^{(2)}$ is a scalar curvature
on the string world sheet and $c$ is an undetermined constant.
The second term comes from the integration of gauge fields
$A$ and $\bar A$ : we have evaluated $\det^{-1}(\cosh\,r+1)$
by using heat kernel method.
The obtained action \tviii\ can be regarded as the action of
string $\sigma$
model whose target space metric $g_{\mu\nu}$ and dilaton field
$\Phi$ are given by,
$$g_{rr}=1, \ \ g_{\phi \phi }=\tanh^2 r, \ \ g_{r\phi }
=g_{\phi r}=0\ , \eqn\tix$$
$$\Phi=2\,\ln (\cosh\, r+1)+c \ . \eqn\tx$$
When $r$ goes to infinity, the action \tviii\ takes the form:
$$I={k \over \pi}\int d^2z\partial r\bar\partial r
-{2 \over \pi}\int d^2z \sqrt h R^{(2)}r
+{k \over \pi}\int d^2z \partial \phi \bar\partial \phi
\ .\eqn\txi$$
The system described by the action \txi\  is
a direct product of Coulomb gas $r$ and free boson $\phi$.
On the other hand, when $r$ vanishes, $g_{\phi\phi}$ in Eq.\tix\
vanishes and
the target space metric \tix\ becomes singular. However the target space
curvature $R$, which is given by
$$R={4 \over \cosh^4 r}\ , \eqn\txii$$
is not singular ($R \rightarrow 4$ when $r \rightarrow 0$).
Therefore we find that the target space metric has the form of
semi-infinite cigar.

The theory corresponding to the case that we gauge the axial symmetry
generated by {\it non-}compact abelian subgroup of $SU(1,1)$
is given by replacing $\sigma_2$ by $i\sigma_3$ in the parametrization
\tiii\ or simply analytic continuating $\phi \rightarrow it$.
The corresponding target space metric is given by,
$$g_{rr}=1, \ \ g_{tt}=-\tanh^2 r, \ \ g_{rt}
=g_{tr}=0\ , \eqn\txiii$$
If we define new coordinates $u$, $v$ by imitating the Kruskal
coordinates of the Schwarzschild black hole in four dimensions
$$u=-\sinh\,r\,e^{-t}\ , \ \ \ v=\sinh\,r\,e^t \ , \eqn\txiv$$
the target space metric can be rewritten by
$$g_{uu}=g_{vv}=0, \ \ g_{uv}=g_{vu}=-{1 \over 1-uv}\ . \eqn\txv$$
This metric \txv\ tells that the surface which is defined by $uv=0$
($r=0$) is a horizon, which is a null surface\foot{Null surfaces
in general $D$ dimensional space are defined as
follows: Let the equation $f(x_i)=0$
define $D-1$ dimensional hypersurface embedded in $D$ dimensional space.
Here $x_i$ ($i=1,\cdots,D$) are coordinates in $D$ dimensional
space. Then a normal vector ${\bf F}$ is given by
${\bf F}=(\partial_1 f, \partial_2 f, \cdots, \partial_D f)$.
The null surface is a surface
where the normal vector ${\bf F}$ has the vanishing length,
${\bf F}\cdot {\bf F}=g^{ij}\partial_if\partial_jf=0$.
In the case that we are now considering, the function $f$ is given by
$f=uv$ and we find
${\bf F}\cdot {\bf F}=-4(1-uv)uv$. Since ${\bf F}\cdot {\bf F}$
vanishes when $uv=0$, the surface defined by the equation $uv=0$
is a null surface.}
without curvature
singularities. Therefore the target space can be regarded as
a two-dimensional black hole.
On the other hand, the surface which satisfies $uv=1$
is a really singular surface where the scalar curvature diverges.
The target space is divided into six regions:
\item{{\rm I,\ IV}} : $uv<0$ ($u>0$, $v<0$ in the region I and
$u>0$, $v<0$ in the region IV)
\item{{\rm II,\ III}} : $0<uv<1$ ($u,v>0$ in the region II and
$u,v<0$ in the region III)
\item{{\rm V,\ VI}} : $uv>1$ ($u,v>0$ in the region V and
$u,v<0$ in the region VI)

\noindent
The meanings of the regions V and VI, which lie beyond the singularity,
will be clarified in the next section. The regions V and VI are related
to the regions I and IV by the duality.

Finally we consider the case that the vector symmetry generated
by compact or non-compact abelian subgroup of $SU(1,1)$ is gauged.
By using the parametrization \tiii\ and the following gauge fixing
condition, instead of \tvii ,
$$\phi_L=\phi_R=\phi \ ,\eqn\txvi$$
we obtain the following gauge fixed action,
$$I={k \over \pi}\int d^2z(\partial r\bar\partial r \pm \coth^2r
\partial \phi \bar\partial \phi)
+({\rm dilaton\ term})\ .\eqn\txvii$$
Here the sign $+$ ($-$) corresponds to the case that
the compact (non-compact) subgroup is
gauged.
The obtained metric
$$g_{rr}=1, \ \ g_{\phi\phi}=\pm\coth^2 r, \ \ g_{r\phi}
=g_{\phi r}=0\ , \eqn\txviii$$
is singular, \ie , $g_{\phi\phi}$ diverges at $r=0$.
The target space has the form like a trumpet or funnel.
The vector transformation $g \rightarrow h^{-1}gh$
has a fixed point $r=0$ in the parametrization \tiii ,
$g\rightarrow h^{-1}gh=g$. Therefore the singularity appears
in a way similar to orbifolds.
On the other hand, the axial transformation
$g \rightarrow hgh$ has no fixed point and we obtain a
smooth manifold.

\chapter{The duality of the target space manifold}

In this section we discuss the duality structure of the target space.
We also consider the more general manifolds which have
Killing symmetries.
More detailed discussion was given in Ref.\rpi . (See also Refs. \rpii ,
\rpiii\ and references therein.)

As shown in the last section, the target space when we gauge
the axial symmetry generated
by non-compact abelian subgroup is divided by six regions.
If we define new coordinates $(t,s)$ in regions V and VI where $uv>1$,
$$u=\cosh\,s\,e^{-t}\ , \ \ \ v=\cosh\,s\,e^t \ , \eqn\txix$$
the target space metric can be rewritten by
$$g_{ss}=1, \ \ g_{tt}=-\coth^2s, \ \ g_{st}=g_{ts}=0\ . \eqn\txx$$
The obtained metric \txx\ is identical with the metric \txviii\ which
appears when the vector symmetry generated
by non-compact abelian subgroup is gauged. This imply the regions
V and VI are equivalent to the region I and IV since
the corresponding conformal field theories are equivalent.
Kiritsis has proved the equivalence as path integrals of GWZW
model\pref\rrvii
The equivalence is a generalization of the invariance which appears
in case of a boson compactified on the circle.
When we change the radius $R$ of
the circle into $1/R$, the spectrum of the boson is invariant.

Before we gauge the abelian symmetry, the WZW action \ti\ has
axial $\otimes$ vector abelian symmetries. After we gauge
the axial (vector) symmetry, there remains a global vector (axial)
symmetry. The global symmetry becomes a Killing symmetry of the
target space metric. As we show later, Killing symmetries connect
the metrics between two different target spaces.
We define Killing symmetries as follows : Let's consider the metric
$g_{\mu\nu}=g_{\mu\nu}(x_0,x_1,\cdots,x_{D-1})$
in $D$ dimensional space. Here $x_i$'s are coordinates in the space.
If the metric $g_{\mu\nu}$ does not depend on one of coordinates,
say $x_i$, when we choose a special coordinates system,
$${\partial g_{\mu\nu} \over \partial x_i}=0 \ , \eqn\txxi$$
the metric $g_{\mu\nu}$ has an abelian symmetry,
$$\eqalign{g_{\mu\nu}(x_0,x_1,&\cdots,x_i,\cdots,x_{D-1}) \cr
&\rightarrow g_{\mu\nu}(x_0,x_1,\cdots,x_i+a_i,\cdots,x_{D-1}) \cr
&\ \ \ \ =g_{\mu\nu}(x_0,x_1,\cdots,x_i,\cdots,x_{D-1}) \ ,}\eqn\txxii$$
{\it i.e.}, the metric $g_{\mu\nu}$ is invariant under the
transformation $x_i\rightarrow x_i+a_i$.
Here $a_i$ is a parameter of the transformation.
We call this symmetry as a Killing symmetry.
In case of $SU(1,1)/U(1)$ coset model, the $SU(1,1)$ WZW model \ti\
has global axial symmetry $\delta \phi_L=\delta \phi_R=\alpha $
and global vector symmetry $\delta \phi_L=-\delta \phi_R=\beta $
in the parametrization \tiii . When we gauge the axial symmetry and
fix the gauge symmetry by the gauge condition \tvii , the system has
a global vector symmetry $\delta \phi =\beta $,
which is a Killing symmetry of the target space metric \tix .
Note that the metric in Eq.\tix\ does not depend on $\phi$.

In the following, we show that Killing symmetries connect the metrics
between the two different target spaces. We consider the following
$\sigma $ model action which has $D$ dimensional target space.
$$S=-{1 \over 2}\int d^2z
\tilde g_{\mu\nu}\partial x^\mu \bar\partial x^\nu
+\int d^2z\sqrt h R^{(2)}\Phi + \cdots \ \ \ \ \ \mu,\nu=0,1,\cdots,D-1
\eqn\txxiii$$
Here $\tilde g_{\mu\nu}$ is a sum of $D$ dimensional metric
$g_{\mu\nu}$ and anti-symmetric tensor $a_{\mu\nu}$ :
$$\tilde g_{\mu\nu}= g_{\mu\nu}+a_{\mu\nu} \ . \eqn\txxiv$$
We now assume that $\tilde g_{\mu\nu}$, $\Phi$ and \etc\ do not depend
on one of coordinates, say $x_0$.

The partition function is given by
$$\eqalign{Z=&\int[dx^\mu]e^S \cr
=&\int[dx^\mu]e^{-{1 \over 2}\int d^2z
(\tilde g_{00}\partial x^0 \bar\partial x^0
+\tilde g_{0i}\partial x^0 \bar\partial x^i
+\tilde g_{i0}\partial x^i \bar\partial x^0
+\tilde g_{ij}\partial x^i \bar\partial x^j)
+\int d^2z\sqrt h R^{(2)}\Phi + \cdots } \cr
&\ \ \ \ \ i,j=1,2,\cdots,D-1 \ .}
\eqn\txxiv$$
By using the following formula
$$e^{-ab^2}={1 \over 2\sqrt{\pi a}}\int_{-\infty }^\infty dy
e^{-{y^2 \over 4a}+iby} \ , \eqn\txxv$$
we can rewrite the partition function \txxiv
$$\eqalign{Z=&\int[dx^\mu][dy][d\bar y]\det^{-1}\tilde g_{00}\, \cr
&e^{-{1 \over 2}\int d^2z
\{{y\bar y \over \tilde g_{00}}
+\partial x^0(2i\bar y+\tilde g_{0i} \bar\partial x^i)
+(2iy+\tilde g_{i0}\partial x^i) \bar\partial x^0
+\tilde g_{ij}\partial x^i \bar\partial x^j\}
+\int d^2z\sqrt h R^{(2)}\Phi + \cdots }\ .}
\eqn\txxvi$$
The integration of $x_0$ gives a constraint
$$\partial (i\bar y+\tilde g_{0i} \bar\partial x^i)
+\bar\partial (iy+\tilde g_{i0}\partial x^i)=0 \ , \eqn\txxvii$$
which is solved by
$$i\bar y+\tilde g_{0i} \bar\partial x^i=\bar\partial \phi \ , \ \ \
iy+\tilde g_{i0}\partial x^i=-\partial \phi \ . \eqn\txxviii$$
Here $\phi$ is a real bosonic field.\foot{
This reality assignment is correct when we Wick-rerotate into Minkowski
path integral where $iy$ and $i\bar y$ are real.}
Note that the Jacobian is c-number
when we change the variables $y$ and $\bar y$ into $\phi$.
Furthermore, we can evaluate $\det^-1\tilde g_{00}$
by using heat kernel method.
Then the partition function $Z$ in Eq.\txxiv\ or \txxvi\ is
rewritten by
$$\eqalign{Z&=\int[dx^\mu][d\phi] \cr
&\times e^{-{1 \over 2}\int d^2z
\{ {1 \over \tilde g_{00}}\partial \phi \bar\partial \phi
-{\tilde g_{0i} \over \tilde g_{00}}\partial\phi\bar\partial x^i
+{\tilde g_{i0} \over \tilde g_{00}}\partial x^i \bar\partial \phi
+(\tilde g_{ij}-{\tilde g_{i0}\tilde g_{0j} \over \tilde g_{00}})
\partial x^i \bar\partial x^j\}
+\int d^2z\sqrt h R^{(2)}(\Phi + \ln g_{00})+ \cdots } \ .}
\eqn\txxix$$
If we regard $\phi$ as a new coordinate $x_0'$ of
the target space, $x_0'=\phi$,
we find new metric $\tilde g'_{\mu\nu}$ and new dilaton field $\Phi'$
are given by
$$\tilde g'_{00}={1 \over \tilde g_{00}} \ , \ \
\tilde g'_{0i}=-{\tilde g_{0i} \over \tilde g_{00}}\ , \ \
\tilde g'_{i0}={\tilde g_{i0} \over \tilde g_{00}}\ ,\ \
\tilde g'_{ij}=\tilde g_{ij}
-{\tilde g_{i0}\tilde g_{0j} \over \tilde g_{00}} ,
\eqn\txxx$$
$$\Phi'= \Phi + \ln\,g_{00}\ . \eqn\txxxi$$
Equations \txxx\ and \txxxi\ appeared in Ref.\rpi .
These equations give a generalization of \lq\lq $R \rightarrow 1/R$''
symmetry of a compactified boson.

\chapter{${SU(1,1) \over U(1)}$ coset model and two dimensional gravity}

In this section, we explain the relation between $SU(1,1)/U(1)$
GWZW model and two dimensional gravity.

The covariantly gauge fixed theory of two dimensional gravity
coupled with conformal matter has conformal symmetry.
The action was considered to be
given by a sum of the actions of Faddev-Popov ghost, the conformal
matter and Coulomb gas which corresponds to
Liouville mode $\varphi$ of the metric\pref{\rraii , \rraiii}
The total central charge of this system vanishes.
When the conformal anomaly $c$ of the conformal matter is less than
one, the action successfully describes the two dimensional gravity.
When $c=1$, however, the analysis based on matrix models
has clarified that there exist logarithmic corrections for the
scaling\pref{\rravi , \rravii , \rraviii}
The corrections can be understood by the cut off for the Liouville
mode $\varphi$ : $\varphi \geq 0$\pref{\rraiv , \rrav}
This cut off is necessary in order to
obtain finite theories.

The conformal field theory which describes two dimensional gravity
coupled with $c=1$ conformal matter, \ie , free boson
($d=1$ non-critical string) will satisfy
the following two requirement :
\item{1.} In some asymptotic region,
the theory describes the system of Coulomb gas coupled with
a free boson.
\item{2.} There is a natural cut off for the mode of the
Coulomb gas.

\noindent
The $SU(1,1)/U(1)$ GWZW model \tviii\ corresponding to the case
that the axial symmetry generated
by compact abelian subgroup of $SU(1,1)$ is gauged satisfies
these two requirement if we identify $r$ with the Liouville mode
$\varphi$ and $\phi$ with the free boson.
The one loop analysis of the $\beta $ function
in two dimensional target space $\sigma $ model\refmark\rrai
suggests that the black hole solution \tviii\ is a unique solution
whose $\beta $ functions vanish when there is no
tachyon background (cosmological term).\foot{Two loop analysis
was given in Ref.\rrvii . The effects of the tachyon background
was considered in Ref.\rrxviii .}
Several authors\refmark{\rrxv , \rrxvii , \rrrv}
investigated the relation between $SU(1,1)/U(1)$
coset model and $d=1$ non-critical string.
Especially, the spectrum of $SU(1,1)/U(1)$
coset model was analyzed in detail in Ref.\rrrv .
It was found that the spectrum contains some of the states found in the
studies of $d=1$ non-critical string theory but, in addition,
there appears new states not previously found. The actual spectrum
could be smaller if there is any truncation consistent with
modular invariance.

\chapter{Superstring in Two Dimensional Black Hole}

In this section, superstring theory in two dimensional black hole
background is discussed based on supersymmetric $SU(1,1)/U(1)$ GWZW
model\pref\rrxix

The action $S^{\rm WZW}(G)$ of $N=1$ supersymmetric
WZW model is given by\cref\rix
$$\eqalign{S^{\rm WZW}(G)
=&{k \over 2\pi}\int d^2z d^2\theta {\rm tr}G^{-1}DGG^{-1}\bar DG
\cr
&-{k \over 2\pi}\int dt d^2z d^2\theta
[{\rm tr}G^{-1}DGG^{-1}\bar DG G^{-1}\partial_t G
 +(D \leftrightarrow \bar D {\rm term})] \ . }\eqn\i$$
Here we define
covariant derivatives $D$ and $\bar D$
by using holomorphic and anti-holomorphic Grassmann
coordinates $\theta$  and $\bar \theta$
$$D\equiv {\partial \over \partial \theta}-
\theta{\partial \over \partial z} \ , \ \ \ \
\bar D\equiv {\partial \over \partial \bar \theta}-
\bar \theta{\partial \over \partial \bar z}\ . \eqn\ii$$
The matrix superfield $G$, which is an element of a group \G,
is given by
$$G=\exp (i\sum_a T^a \Phi^a)\ .\eqn\iii$$
Here $\Phi^a$ is a superfield $\Phi^a=\phi^a+\theta\psi^a
+\bar\theta\,\bar\psi^a+\theta\bar\theta f^a$ and $T^a$ is a
generator of the algebra corresponding to \G.
The action \i\ satisfies Polyakov-Wiegmann type formula:
$$S^{\rm WZW}(GH)=S^{\rm WZW}(G)+S^{\rm WZW}(H)+{k \over \pi}
\int d^2z d^2\theta\,{\rm tr}\,G^{-1}DG \bar DH H^{-1}\ .\eqn\iv$$
Here $H$ is also an element of \G.
This formula guarantees that the system described by the action
\i\ has
super Kac-Moody symmetry and $N=1$ superconformal symmetry.

If \F\ is an abelian subgroup of \G, we can gauge
the following global axial symmetry in the action \i ,
which is given by an element $F$ of \F,
$$G \rightarrow FGF\ . \eqn\biii$$
The action of ${\cal G}/{\cal F}$ SGWZW is given by,
$$\eqalign{S^{{\cal G}/{\cal F}}(G, A)=&S^{\rm WZW}(F_LGF_R)-
S^{\rm WZW}(F_L^{-1}F^R)\cr
=&S^{\rm WZW}(G) \cr
&+{k \over 2\pi}\int d^2z d^2\theta\,{\rm tr}(A\bar A+AG\bar A G^{-1}+
G^{-1}DG\bar A+A \bar DG G^{-1}) \ .} \eqn\rrrrv$$
Here $F_{L,R}\in {\cal F}$ and gauge fields $A$, $\bar A$ are defined by,
$$A=F_L^{-1}DF_L , \ \ \ \ \bar A=F_R^{-1}\bar DF_R\ . \eqn\vi$$
The action \rrrrv\ is invariant under the following $U(1)$ gauge transformation
$$\eqalign{G \rightarrow FGF , \ \ A \rightarrow A&+F^{-1}DF ,
\ \ \bar A \rightarrow \bar A+ F^{-1}\bar DF \ . \cr
&(F_{L,R} \rightarrow F_{L,R}F) }\eqn\biv$$

A supersymmetric extension of string theory in a two dimensional
black hole
background is given by setting ${\cal G}=SU(1,1)$ in the action \rrrrv .
We start with considering $SU(1,1)$ SWZW model.
By parametrizing $G$ by,
$$G=\exp ({i \over 2}\Phi_L\sigma_2) \exp ({1 \over 2}R\sigma_1)
\exp ({i \over 2}\Phi_R\sigma_2) , \eqn\vii$$
with $\sigma_i$ the Pauli matrices,
we obtain the action $S^{SU(1,1)}$ of $SU(1,1)$ SWZW model
$$\eqalign{S^{SU(1,1)}=&{k \over 2\pi}\int d^2z d^2\theta
[-{1 \over 2}D\Phi_L\bar D\Phi_L
-{1 \over 2}D\Phi_R\bar D\Phi_R \cr
&-{\rm cosh}\,R\,D\Phi_L\bar D\Phi_R+{1 \over 2}DR\bar DR]\ .}\eqn\ai$$
The holomorphic (anti-holomorphic) conserved currents $J_i$ ($\bar J_i$)
of this system are given by,
$$2kG^{-1}DG=J_1 \sigma_1 +iJ_2 \sigma_2 +J_3 \sigma_3\ , \ \
2kG^{-1}\bar DG=\bar J_1 \sigma_1 +i\bar J_2 \sigma_2
+\bar J_3 \sigma_3\ , \eqn\xxv$$
$$J_i=j_i + \theta{\tilde J_i}+\cdots \ ,\ \ \
\bar J_i=\bar j_i + \bar \theta \bar{\tilde J_i}+ \cdots \ . \eqn\xxvi$$
Here $\cdots$ express the terms which vanish by using
the equations of motion.
If we define new currents $\hat J_i$ and $\bar{\hat J}_i$ by the following
equation
$$\hat J_i=\tilde J_i-{1 \over 2k}\epsilon_{ilm}j_lj_m , \ \
\bar{\hat J}_i=\bar{\tilde J}_i
-{1 \over 2k}\epsilon_{ilm}\bar j_l\bar j_m , \eqn\aii$$
These currents $\hat J_i$ and $\bar{\hat J}_i$ do not depend on the fermion
currents $j_i$ and $\bar j_i$.

By expanding superfields $\Phi_{L,R}$ and $R$ into components,
$$\eqalign{
\Phi_{L,R}&=\phi_{L,R}+\theta\psi_{L,R}+\bar\theta\,\bar\psi_{L,R}+
\theta\bar\theta f_{L,R} , \cr
{R \over 2}&=s+\theta\eta+\bar\theta\bar\eta+\theta\bar\theta g ,}
\eqn\bx$$
we can rewrite the $SU(1,1)$ SWZW action $S^{SU(1,1)}$ in
Eq.\ai\ by a sum of bosonic
$SU(1,1)$ WZW action $\tilde S^{SU(1,1)}$
and free fermion actions:
$$\eqalign{S^{SU(1,1)} =&\tilde S^{SU(1,1)}
+{1 \over 4k\pi}\int d^2z [j_+\bar\partial j_--j_2\bar\partial j_2
+\bar j_+\partial \bar j_--\bar j_2\partial \bar j_2]\ , }\eqn\aiii$$
$$\eqalign{\tilde S^{SU(1,1)}=&{k \over 2\pi}
\int d^2z [-{1 \over 2}(\partial\phi_L\bar\partial\phi_L
+\partial\phi_R\bar\partial\phi_R)\cr
&-{\rm cosh}(2s) \partial\phi_L\bar\partial\phi_R +2\partial s\bar\partial s]
\ .}
\eqn\xxviii$$
Here $j_\pm$ and $\bar j_\pm$ are defined by
$$j_\pm\equiv j_1\pm ij_3\ , \ \ \ \bar j_\pm\equiv \bar j_1\pm i\bar j_3
\ .\eqn\xxix$$
The conserved currents corresponding to the non-supersymmetric
$SU(1,1)$ WZW action $\tilde S^{SU(1,1)}$ \xxviii\ are given by
$\hat J_i$ and $\bar{\hat J}_i$ in Eq.\aii .

Fermionic currents $j_\pm$ and $\bar j_\pm$ can be written as
$$\eqalign{j_\pm =&{k \over2}
\exp (\mp i \phi_R) (\eta\pm{i \over 2}{\rm sinh}(2s)\psi_L)\  ,
\cr
\bar j_\pm =&{k \over 2}\exp (\mp i \phi_L)
(\bar\eta\pm{i \over 2}{\rm sinh}(2s)\bar\psi_R) \ .} \eqn\bi$$
Note that there appear bosonic factors
$\exp (\mp i \phi_R) $ and $\exp (\mp i \phi_L) $.
Due to these factors, the boundary conditions of
$j_\pm$ and $\bar j_\pm$ are twisted although fermions $\eta$,
$\bar\eta$, $\psi_L$ and $\bar\psi_R$, which will be identified later
with space-time fermionic coordinates, should be
periodic or anti-periodic. Therefore the eigenvalues of the zero modes
of fermion number currents $K$ and $\bar K$,
$$K={1 \over 4k}(j_+j_--j_-j_+)\ , \ \ \
\bar K={1 \over 4k}(\bar j_+\bar j_--\bar j_-\bar j_+)\ , \eqn\xxx$$
which satisfy the following operator product expansions
$$K(z)j_\pm(w)\sim\pm{1 \over z-w}j_\pm\ , \ \ \
\bar K(\bar z)\bar j_\pm(\bar w)\sim\pm{1 \over \bar z-\bar w}
\bar j_\pm\ , \eqn\xxxi$$
are not quantized.

We now gauge the abelian symmetry in the action \ai\ by following
Eq. \rrrrv .
We consider the case that the abelian symmetry is generated by
$\sigma_2$. Since the abelian symmetry is compact, the resulting theory
describes the Euclidean black hole.
The theory of the Lorentzian black hole can be obtained by replacing
$\sigma_2$ by $i\sigma_3$ or simply by analytic continuating
$\Phi_{L,R}\rightarrow i\Phi_{L,R}$.

By using the parametrization \vii ,
the $SU(1,1)/U(1)$ gauged SWZW action
takes the form
$$\eqalign{S^{SU(1,1)/U(1)}=&S^{SU(1,1)}
+{k \over 2\pi}\int d^2z d^2\theta [4(1+{\rm cosh}\,R)A\bar A \cr
&+2iA(\bar D \Phi_L+{\rm cosh}\,R\,\bar D \Phi_R)
+2i(D \Phi_L+{\rm cosh}\,R\, D \Phi_R)\bar A] \ .}\eqn\viii$$
Here $S^{SU(1,1)}$ is $SU(1,1)$ SWZW action in Eq.\ai .
By the following redefinitions,
$$\eqalign{\Phi\equiv &\Phi_L-\Phi_R\ , \cr
A'\equiv &A+{i \over 2}{{\rm cosh}\,R\,D\Phi_L+D\Phi_R \over 1
+ {\rm cosh}\,R}\ , \cr
\bar A'\equiv &\bar A
+{i \over 2}{{\rm cosh}\,R\,\bar D\Phi_R+\bar D\Phi_L \over 1
+ {\rm cosh}\,R}\ , }
\eqn\ix$$
the action \viii\ can be rewritten as follows,
$$\eqalign{S^{SU(1,1)/U(1)}={k \over 2\pi}&\int d^2z d^2\theta
[{1 \over 2}{\rm tanh}^2{R \over 2}\,D\Phi\bar D\Phi \cr
&+{1 \over 2}DR\bar DR+4(1+{\rm cosh}R)\,A'\bar A' \ . }\eqn\x$$
The action \viii\ and \x\ are invariant under the following
infinitesimal gauge transformation corresponding to Eq.\biv ,
$$\delta \Phi_L=\delta \Phi_R=\Lambda \ , \ \
\delta A=-{i \over 2}D\Lambda \ , \ \
\delta \bar A=-{i \over 2}\bar D\Lambda \ . \eqn\xi$$
We fix this gauge symmetry by imposing the following gauge condition
$$\Phi_L=-\Phi_R=\tilde \Phi \ . \eqn\xii$$
By integrating gauge fields $A$ and $\bar A$ in
the action \viii\ or \x ,\foot{
The integration of the gauge fields induces the dilaton term in the action
but we now neglect this term. The gauge fixed action which is correct
at the quantum level is given later in this paper.}
and by integrating auxiliary fields,
we obtain the following action,
$$\eqalign{
S^{(1)}={k \over \pi}&\int d^2 z
[{\rm tanh}^2 s (\partial \phi \bar \partial \phi
- \partial \bar \psi \,\bar \psi
+\psi \bar \partial \psi) \cr
&-2{{\rm sinh}\,s \over {\rm cosh}^3 s}(\eta\psi\bar\partial \phi
+\bar\eta\bar\psi\partial \phi)
+4\,{\rm tanh}^2 s\, \eta\bar\eta\psi\bar\psi \cr
&+\partial s \bar \partial s - \partial \bar \eta \,\bar \eta
+\eta \bar \partial \eta] \ .}\eqn\xiii$$
Here we write superfields $\tilde\Phi$ and $R$ in terms of components:
$$\eqalign{
\tilde\Phi&=\phi+\theta\psi+\bar\theta\,\bar\psi+\theta\bar\theta f
\ , \cr
{R \over 2}&=s+\theta\eta+\bar\theta\bar\eta+\theta\bar\theta g\ .}
\eqn\xii$$
This system has $N=1$ supersymmetry since the starting action \viii\
and gauge condition \xi\ are manifestly supersymmetric.
In fact, this action is nothing but the action of
(1,1) supersymmetric $\sigma$ model\refmark\rxv in two dimensional
black hole background.

The $N=1$ supersymmetry in the action \xiii\
is extended to $N=2$ supersymmetry since this action is invariant
under the following holomorphic
(anti-holomorphic) $U(1)$ symmetry:
$$\eqalign{\delta\psi =&-{u(z) \over {\rm tanh}s} \eta \cr
\delta\eta =&u(z) {\rm tanh}s \,\psi \ ,}\eqn\xv$$
and
$$\eqalign{\delta\bar\psi =&-{\bar u(\bar z) \over {\rm tanh}s} \bar
\eta \cr
\delta\bar\eta =&\bar u(\bar z) {\rm tanh}s \,\bar\psi\ .}\eqn\xvi$$
Here $u(z)$ ($\bar u(\bar z)$) is a holomorphic (anti-holomorphic)
parameters of the transformation.
The transformations \xv\ and \xvi\ tell that the currents of
this $U(1)$ symmetry
can be regarded as fermion number currents
with respect to
space-time fermion coordinates, $\eta$, $\psi$,
$\bar\eta$ and $\bar\psi$.
By commuting this $U(1)$ symmetry transformation with the original $N=1$
supersymmetry transformation, we obtain another supersymmetry
transformation
and we find that the action has $N=2$ supersymmetry. On the other hand,
in case of the Lorentzian black hole, the obtained algebra is not
exactly $N=2$ superconformal algebra.\foot{
Note that any Lorentzian manifold is not K\"ahler.}
Usual $N=2$ superconformal algebra is given by
$$\eqalign{\{G_n^+, G_m^-\}=&4L_{n+m}+2(m-n)J_{n+m}+{c \over 12}m(m^2-1)
\delta_{m+n,0} \ , \cr
[J_n, G_m^\pm]=&\pm G_{n+m}\ , \ \ \ \ {\rm etc.}}
\eqn\xvii$$
and the hermiticities of the operators are assigned by
$$(G_n^+)^\dagger =G_{-n}^-\ , \ \ \ J_n^\dagger =J_{-n}\ . \eqn\xviii$$
The algebra which appears in the Lorentzian case is identical with Eq.\xvii ,
but the assignment of the hermiticities is different from Eq.\xviii :
$$(G_n^+)^\dagger =G_{-n}^+\ , \ \ \ (G_n^-)^\dagger =G_{-n}^-\ , \ \ \
J_n^\dagger =-J_{-n} \ . \eqn\xix$$
This is not so surprising since this algebra also appears in flat two
dimensional Lorentzian space-time which is a subspace of
flat ten dimensional space-time in usual Neveu-Schwarz-Ramond model.
Even in the Lorentzian case, we have a $U(1)$ current
and superstring theories can be constructed by imposing
GSO projection.

In order to consider the spectrum of this theory, we choose the following
gauge condition instead of Eq. \xi\ ,
$$\bar DA-D\bar A=0 \ . \eqn\xx$$
This gauge condition allows us to parametrize the gauge fields $A$ and $\bar A$
as
$$A=D\Pi \ , \ \ \ \bar A=-\bar D\Pi \ . \eqn\xxi$$
By shifting the fields $\Phi_{L,R}$,
$$\Phi_L \rightarrow \Phi_L +2i\Pi \ ,
\ \ \Phi_R \rightarrow \Phi_R -2i\Pi \ ,
\eqn\xxii$$
the gauge fixed action $S^{(2)}$
is given by a sum of $SU(1,1)$ SWZW action $S^{SU(1,1)}$ in
Eq. \ai , free field action $S^{\Pi}$
and (free) ghost action $S^{\rm FP}$.
$$\eqalign{S^{(2)}=&S^{SU(1,1)}+S^{\Pi}+S^{\rm FP}\ , \cr
S^{\Pi}=&-{4k \over \pi}\int d^2z d^2\theta D\Pi \bar D\Pi\ , \cr
S^{\rm FP}=&{k \over 2\pi}\int d^2z d^2\theta BD \bar DC\ . }\eqn\xxiii$$
Here $B$ and $C$ are anti-ghost and ghost superfields.

The BRS charge
$Q_{\rm B}$
which defines the physical states is given by
$$Q_{\rm B}=\oint dz C(D\Pi-{i \over 4k}J_2)+\oint d\bar z
C(\bar D\Pi+{i \over 4k}\bar J_2) \ . \eqn\xxiv$$
This BRS charge gives constraints on the physical states,
$$D\Pi-{i \over 2}J_2=\bar D\Pi+{i \over 2}\bar J_2=0 \ , \eqn\xxvii$$
which tell that $B$, $C$, $\Pi$ and $J_2$ (or $\bar J_2$) make
so-called \lq\lq quartet" structure\refmark\rxv
similar to the structure which appeared
in the quantization of Neveu-Schwarz-Ramond model based on BRS
symmetry\pref{\rxvi , \rxvii}

The action which describes superstring theory in the two dimensional
black hole is simply given by a sum of $SU(1,1)$ WZW action \xxviii ,
free fermion actions \aiii\ and the actions of free superfield
and free ghost and anti-ghost superfields \xxiii . Furthermore
the constraints \xxvii\ imposed by th BRS charge \xxiv\ can be easily
solved with respect to free superfields $\Pi$. Therefore if we can find
the spectrum of the bosonic string in the two dimensional
black hole,\refmark{\ri , \rii} we can also find the spectrum
of this string theory.

The $U(1)$ current, which corresponds to the transformations \xv\
and \xvi\ are given by\cref\rv
$$J={-2i \over k-2}\hat J_2+{k \over k-2}K\ , \ \ \ \bar J
={-2i \over k-2}\bar{\hat J}_2+{k \over k-2}\bar K\ . \eqn\xxxii$$
Here $\hat J_2$ and $\bar{\hat J}_2$ are defined by Eq.\aii\ and
fermion number currents $K$ and $\bar K$ are defined by
Eq.\xxx .
These $U(1)$ currents commute with the BRS charge \xxiv\ and
we can impose GSO projection consistently.
Note that GSO projection does not give any constraint on the
representations of $SU(1,1)$ current algebra since
the eigenvalues of the zero modes in the currents $K$ and $\bar K$ are
not quantized although those in $J$ and $\bar J$ are quantized.

\chapter{Summary and Discussion}

We have reviewed the recently proposed
string theory in two dimensional black hole background.
Especially, the structure of the duality in the
target space was discussed. Furthermore we analyzed the duality
structure of more general target space
manifolds which have Killing symmetries.
The duality is a generalization of \lq\lq $R \rightarrow 1/R$" symmetry
of a compactified boson.
We have derived a formula which appeared in Ref.\rpi .
This formula connects two different
manifolds which are dual to each other.

We have also discussed supersymmetric $SU(1,1)/U(1)$ gauged
Wess-Zumino-Witten model.
Due to Kazama-Suzuki\refmark\riv mechanism,
this model has $N=2$ superconformal symmetry.
When the central charge $c=15$,
this conformal field theory describes a critical
Neveu-Schwarz-Ramond string theory.
The $N=1$ supergravity coupled with $c={3 \over 2}$
($\hat c=1$) superconformal matter would be described by
this critical theory.
The $N=2$ superconformal symmetry of this model suggests
that pure $N=2$ supergravity would be also described by this model
when $c=6$ ($k=4$).
Due to $N=2$ superconformal symmetry, superstring theories in the two
dimensional black hole background can be constructed by imposing GSO
projection.
We expect that this superstring theory
would be equivalent to the matrix models which have space-time
supersymmetry\refmark{\rvi, \rvii}
and topological superstring theories based on $N=2$
superconformal topological field theories\pref\rviii

Recently string models based on ${SU(1,1)\times U(1) \over U(1)}$
coset model were considered in Refs.\rxiii\ and \rxiv .
These models describe
the strings in two\refmark\rxiii or three\refmark\rxiv dimensional
charged black holes. By adjusting the radius of the $U(1)$ boson,
we will obtain $N=2$ superconformal theory with $c>3$\refmark\riii
in the same way as $N=2$ minimal model was constructed from
${SU(2)\times U(1) \over U(1)}$\pref\rx The obtained model should be
equivalent to the model discussed here.

\ack{
I would like to acknowledge discussions with N. Ishibashi, M. Li,
J. Lykken and A. Strominger. I am also indebted to M. Kato, E. Kiritsis,
A. Sugamoto, T. Uchino and S.-K. Yang for the discussion
at the early stage.
I wish to thank the theory groups of SLAC, UC Santa Barbara
and Fermilab, where a part of this work was done.
I am grateful to M. Rocek for drawing my attention to Ref.\rpi .
This work is supported by Soryuushi Shougakkai.}

\endpage
\refout
\bye